\documentclass[twocolumn,prb,amsmath,amssymb,showpacs]{revtex4}

\usepackage{bm}
\usepackage{graphicx}

\def \be{\begin{equation}}
\def \ee{\end{equation}}
\def \bmlett{\begin{mathletters}}
\def \emlett{\end{mathletters}}

\def \ve{\varepsilon}

\def \pd{\phantom{\dagger}}

\def \ra{\rightarrow}

\begin{document}

\bibliographystyle{simpl1}


\title{Noise and Measurement Efficiency of a Partially Coherent Mesoscopic Detector}

\author{A. A. Clerk and A. D. Stone}
\affiliation{ Departments of Applied Physics and Physics, Yale
University, New Haven CT, 06511, USA\\ Jan. 19, 2004}

\begin{abstract}
We study the noise properties and efficiency of a mesoscopic
resonant-level conductor which is used as a quantum detector, in
the regime where transport through the level is only partially
phase coherent.  We contrast models in which detector incoherence
arises from escape to a voltage probe, versus those in which it
arises from a random time-dependent potential.  Particular
attention is paid to the back-action charge noise of the system.
While the average detector current is similar in all models, we
find that its noise properties and measurement efficiency are
sensitive both to the degree of coherence and to the nature of the
dephasing source.  Detector incoherence prevents quantum limited
detection, except in the non-generic case where the source of
dephasing is not associated with extra unobserved information.
This latter case can be realized in a version of the voltage probe
model.
\end{abstract}

\maketitle

\section{Introduction}

Motivated primarily by experiments involving solid-state qubit
systems, attention has recently turned to examining the properties
of mesoscopic conductors viewed as quantum detectors or amplifiers
\cite{Gurvitz,AleinerQPC, Levinson, Buks,Stodolsky, KorotkovQPC,
AverinRLM, MakhlinRMP, DevoretNature, KorotkovSN, Pilgram, Me,
Averin, ShnirmanSmallV}. Of particular interest is the issue of
quantum limited detection-- does a particular detector have the
minimum possible back-action noise allowed by quantum uncertainty
relations? Reaching the quantum limit is crucial to the success of
a number of potential experiments in quantum information physics,
including the detection of coherent qubit oscillations in the
noise of a detector \cite{KorotkovSN}. The measurement efficiency
of a number of specific mesoscopic detectors has been studied
\cite{Gurvitz, KorotkovQPC, AverinRLM, DevoretNature, Pilgram}, as
have general conditions needed for quantum limited detection
\cite{Averin, Me}. Recent studies of a broad class of phase
coherent mesoscopic scattering detectors \cite{Pilgram, Me} have
helped establish a general relation between back-action noise, the
quantum limit and information.  The back-action charge noise of
these detectors was found to be a measure of the total accessible
information generated by the detector's interaction with a qubit,
and the quantum limit condition to imply the lack of any ``wasted"
information in the detector not revealed at its output.

An important unanswered question regards the role of detector
coherence-- does a departure from perfectly coherent transport in
the detector necessarily imply a deviation from the quantum limit?
One might expect that dephasing will have a negative impact, as
there will now be extraneous noise associated with the source of
dephasing. However, if there were unused phase information in the
coherent system, one might expect the addition of dephasing to
bring the detector closer to the quantum limit, as this unused
phase information will be eliminated.  Addressing the influence of
dephasing concretely requires an understanding of its effects on
the noise properties of a detector. In the case where the detector
is a mesoscopic conductor, the influence of dephasing on the
output current noise has received considerable attention
\cite{Buttiker}; in contrast, its influence on the back-action
charge noise has only been addressed in a limited number of cases
\cite{ButtikerNote,Seelig}. Note that a recent experiment by
Sprinzak et. al\cite{Sprinzak} using a point-contact detector
suggests that the back-action noise is independent of dephasing.

To study the role of detector incoherence, we focus here on the
case where the mesoscopic scattering detector is a
non-interacting, single-level resonant tunneling structure, with
the signal of interest (e.g., a qubit) modulating the energy of
the level. This model provides an approximate description of
transport through a quantum dot near a Coulomb blockade charge
degeneracy point, in the limit where the dot has a large level
spacing. Such a system could act as a quantum detector of, e.g., a
double-dot qubit.  The resonant level detector is also
conceptually similar to detectors using the Josephson
quasiparticle (JQP) resonance in a superconducting single electron
transistor \cite{AverinJQP,ChoiJQP,ClerkJQP}, as have been used in
several recent qubit detection experiments \cite{NakamuraCPB,
KonradCPB}. In these systems the resonance is between two
transistor charge states, one of which is broadened by
quasiparticle tunneling, and the signal of interest modulates the
position of the resonance. Despite the incoherence of the
resonance-broadening here, it has been shown theoretically that
one can still make a near quantum-limited measurement using the
JQP process \cite{ClerkJQP}.

The detector properties of a fully coherent resonant level model
were studied comprehensively by Averin in Ref.
\onlinecite{AverinRLM}, who found that detection can be
quantum-limited in the small voltage regime (as follows also from
the general analysis in Refs. \onlinecite{Pilgram} and
\onlinecite{Me}), and near-quantum limited in the large voltage
regime.  We are interested now in how the addition of dephasing
changes these conclusions.  The influence of dephasing on the
noise properties of the resonant level model also has intrinsic
interest, as this is one of the simplest systems with non-trivial
energy-dependent scattering. Standard treatments
\cite{ButtikerVProbe} indicate that the effects of dephasing on
the resonant level cannot be identified via the energy-dependence
of the average current-- both the coherent and incoherent models
yield a Lorentzian form for the conductance.  In contrast, the
noise properties of the coherent and incoherent resonant level
models are significantly different; while this is known for the
current noise, we show that it can also hold for the back-action
charge noise.

To study the effects of dephasing, we will use two general models.
The first corresponds to dephasing due to unobserved escape from
the level, and will be modelled using the voltage probe model
developed by B\"{u}ttiker \cite{ButtikerVProbe}. The second will
correspond to dephasing arising from a random time-dependent
external potential.  This approach is particularly appealing, as
it allows a simple heuristic interpretation of the effects of
dephasing, and allows for a clear separation between pure
dephasing effects and inelastic scattering.

In general, we find that both the magnitude of dephasing and the
nature of the dephasing source need to be determined in order to
evaluate the effect of incoherence on the quantum limit. Dephasing
prevents ideal quantum-limited detection, except in a simple,
physically-realizable version of the voltage probe model. The
latter model is unique, as one can show that there is no extra
unobserved information produced by the addition of dephasing (i.e.
the addition of the voltage probe). We also find that the voltage
probe model can yield very similar noise properties to a
sufficiently ``slow" random potential model in the small voltage
limit; this correspondence however is lost at larger voltages.

\section{Coherent Detector}

We begin by briefly reviewing the properties of a coherent
mesoscopic scattering detector, as discussed in Refs.
\onlinecite{Pilgram} and \onlinecite{Me}.  In the simplest case,
the detector is a phase coherent scattering region coupled to two
reservoirs ($1$ and $2$) via single channel leads, and is
described by the scattering matrix:
\begin{equation}
    s_0(\ve) =  e^{i \alpha_0(\ve)}
    \left(
        \begin{array}{cc}
          e^{i \beta_0(\ve)} \sqrt{1-T_0(\ve)}  & -i   \sqrt{T_0(\ve)}\\
          -i \sqrt{T_0(\ve)} & e^{-i \beta_0(\ve)} \sqrt{1-T_0(\ve)}  \\
        \end{array}
    \right).
    \label{GenS0Defn}
\end{equation}
There are three parameters which determine $s_0$: the overall
scattering phase $\alpha_0(\ve)$, the transmission coefficient
$T_0(\ve)$, and the relative phase between transmission and
reflection, $\beta_0(\ve)$.  The only assumption in Eq.
(\ref{GenS0Defn}) is that time-reversal symmetry holds; as
discussed in Ref. \onlinecite{Me}, the presence or absence of
time-reversal symmetry is irrelevant to reaching the quantum
limit.  Note that if $s_0$ has parity symmetry, the phase
$\beta_0$ is forced to be zero, which implies that there is no
energy-dependent phase difference between transmitted and
reflected currents.

The conductor described by $s_0$ is sensitive to changes in the
potential in the scattering region, and may thus serve as a
detector of charge.  A sufficiently slow input signal $v(t)$ which
produces a weak potential in the scattering region will lead to a
change in average current given by $\delta \langle I(t) \rangle =
\lambda v(t)$, where $\lambda$ is the zero-frequency gain
coefficient of the detector. In the case where the potential
created by the signal (i.e. qubit) is smooth in the scattering
region, and where the drain-source voltage tends to zero (i.e.
$\mu_1 = \mu_2 + eV, eV \ra 0$) the zero frequency noise
correlators of the system are given by \cite{Pilgram, Me}:
\begin{subequations}
\begin{eqnarray}
    \lambda & = &
        \frac{e^3 V}{h} \partial_\ve \big|[s_0(\ve)]_{12} \big|^2 =
        \frac{e^3 V}{h} \partial_\ve T_0(\ve)  \label{CohGain}\\
    S_I & = & \frac{2 e^3 V}{h} T_0(\ve) R_0(\ve)
        \label{CohSI}\\
    S_Q & = & \frac{e^3 V \hbar}{\pi} \times
        \nonumber \\
        && \left(
        \frac{ \left[\partial_\ve T_0(\ve)  \right]^2}
            {4 T_0(\ve) R_0(\ve)  }
        +   T_0(\ve) R_0(\ve) \left[ \partial_\ve
        \beta(\ve)  \right]^2 \right)
             \label{CohSQ}
\end{eqnarray}
\end{subequations}
Here, $S_I$ and $S_Q$ are the zero-frequency output current noise
and back-action charge noise, $R_0 \equiv 1-T_0$ is the reflection
coefficient, and all functions should be evaluated at $\ve = \mu$,
where $\mu$ is the chemical potential of the leads.  Note that the
charge $Q$ here refers to the total charge in the scattering
region, and is not simply the integral of the source-drain current
$I$.  Also note that throughout this paper, we concentrate on the
case of zero temperature.

We are interested in the measurement efficiency ratio $\chi$,
defined as
\begin{equation}
    \chi \equiv \frac{\hbar^2 \lambda^2}{S_I S_Q}.
\end{equation}
In the case of a qubit coupled to the detector, $\chi$ represents
the ratio of the measurement rate to the back-action dephasing
rate in a quantum non-demolition setup \cite{MakhlinRMP,
DevoretNature,Me}, and the maximum signal to noise ratio in a
noise-spectroscopy experiment \cite{KorotkovSN}.  $\chi$ is
rigorously bounded by unity \cite{Averin,Me}, and reaching the
quantum limit corresponds to having $\chi = 1$. If we view our
detector as a linear amplifier, achieving $\chi=1$ is equivalent
to having the minimum possible detector noise energy \cite{Caves,
DevoretNature,Me}.

Even in the $V \ra 0$ limit, the general 1D scattering detector
described above fails to reach the quantum limit because of unused
information available in the phase $\beta(\ve)$:
\begin{equation}
    \chi \equiv \frac{\hbar^2 \lambda^2}{S_I S_Q}
        = \frac{1}{
             1 + \left( 2 T_0 R_0
            \frac{
            \partial_\ve \beta_0}
            {\partial_\ve T_0 }  \right)^2
            }
\end{equation}
As noted, $\beta$ is the energy-dependent relative phase between
reflection and transmission, and in principle is accessible in an
experiment sensitive to interference between transmitted and
reflected currents \cite{Sprinzak}.  The presence of parity
symmetry would force $\beta = 0$, and would thus allow an
arbitrary one-channel phase-coherent scattering detector to reach
the quantum limit in the zero-voltage limit \cite{KorotkovQPC,
Pilgram, Me}.  Note that this discussion neglects screenings
effects; such effects have been included within an RPA scheme in
Ref. \onlinecite{Pilgram}, where it was shown they did not effect
$\chi$.

We now specialize to the case where our scattering region is a
single resonant level.  Taking the tunnel matrix elements to be
independent of energy, the scattering matrix in the absence of
dephasing is determined in the usual way by the retarded Green
function of the level.  Letting $1$ ($2$) denote the $L$ ($R$)
lead, we have:
\begin{equation}
    s_0(\ve) =  \hat{1} - i \sqrt{\Gamma_i \Gamma_j}
    \cdot G^{R}(\ve)
        =
    \hat{1} - \frac{
        i \sqrt{\Gamma_i \Gamma_j} }
        {\ve-\ve_d + i \Gamma_0 / 2}
    \label{S0Defn}
\end{equation}
Here, $\Gamma_L$ ($\Gamma_R$) is the level broadening due to
tunneling to the left (right) lead; $\Gamma_0 = \Gamma_L +
\Gamma_R$ represents the total width of the level due to
tunneling.
The parameters appearing in Eq. (\ref{GenS0Defn}) are given by:
\begin{subequations}
\label{RLMParams}
\begin{eqnarray}
    \alpha_0(\ve) & = & -\arctan\left(\frac{\Gamma_0}{2 (\ve - \ve_d)} \right) \\
    T_0(\ve) & = & \frac{ \Gamma_L \Gamma_R}{(\ve- \ve_d)^2 + \Gamma_0^2 / 4}
        \label{T0Eqn}\\
    \beta_0(\ve) & = & \arctan\left( \frac{\Gamma_R -
    \Gamma_L}{2 (\ve- \ve_d)} \right) \label{BetaEqn}
\end{eqnarray}
\end{subequations}
Note that there is only one non-trivial eigenvalue of $s_0(\ve)$,
given by $e^{2 i \alpha_0}$ (i.e. there is a scattering channel
which decouples from the level).

One finds for the measurement efficiency at zero voltage:
\begin{equation}
    \chi = \frac{ (\ve_d-\mu)^2 }{(\ve_d-\mu)^2 + (\Gamma_L-\Gamma_R)^2/4}
\end{equation}
As per our general discussion above, the coherent resonant level
detector is only quantum limited if there is parity symmetry, i.e.
$\Gamma_L = \Gamma_R$; if this condition is not met, there is
unused information available in the phase $\beta(\ve)$.  For the
resonant level model, the effects of this unused information can
be minimized by working far from resonance (i.e. $|\ve_d-\mu| \gg
0$), as this suppresses the information in the phase $\beta(\ve)$
faster than that in the amplitude $T_0(\ve)$.  It is worth noting
that introducing a third lead to represent dephasing (as we do in
the next section) breaks parity in a similar manner to simply
having $\Gamma_L \neq \Gamma_R$, and its effect may be understood
in similar terms.  Note also that for $\Gamma_L = \Gamma_R, \chi
=1$ for any value of $\mu$. This may seem surprising as the gain
vanishes when $\ve_d = \mu$ (i.e. the peak of the resonance
lineshape). However, the noise also vanishes at this point in just
the manner required to maintain $\chi = 1$.  This feature does not
carry over in any of the models of dephasing; $\chi$ will depend
on the position of $\mu$ and will not be maximized for
$\ve_d=\mu$.

\section{Dephasing From Escape}

\subsection{General setup}

We first treat the effects of dephasing by using the
phenomenological voltage probe model developed by Buttiker
\cite{ButtikerVProbe}. A fictitious third lead is attached to the
resonant level, with its reservoir chosen so that there is no net
current flowing through it. Nonetheless, electrons may enter and
leave this third lead incoherently, leading to a dephasing effect.
We term this dephasing by escape, as it models electrons actually
leaving the level.  In practice, the Green function of the level
is broadened by an additional amount $\Gamma_{\varphi}$ over the
elastic broadening $\Gamma_0$; one also has to explicitly consider
the contribution to the current and noise from electrons which
enter and leave the voltage probe incoherently.

More concretely, assuming that there are $M$ propagating channels
in the voltage probe lead, our detector is now described by a
$(2+M) \times (2+M)$ scattering matrix $s_{big}$.  We assume
throughout the presence of time-reversal symmetry; the results
presented here are independent of this assumption. The
(non-unitary) $2 \times 2$ sub-matrix of $s_{big}$ describing
direct, coherent scattering between the physical leads is given by
\cite{StoneLee}:
\begin{equation}
    \tilde{s}_0(\ve) \equiv s_0(\ve + i \Gamma_{\varphi}/2)
\label{stild}
\end{equation}
where $s_0(\ve)$ is given by Eq. (\ref{S0Defn}).  This matrix
continues to have a decoupled scattering channel (i.e. eigenvalue
1), while the eigenvalue of the coupled channel becomes:
\begin{equation}
    e^{2 i \alpha_0(\ve)}  \ra
        \sqrt{1 - T_{\varphi(\ve)} } e^{2 i \alpha(\ve)}
\end{equation}
with
\begin{eqnarray}
    \tilde{G}^R(\ve) & = &
        \frac{ 1}{\ve - \ve_d + i \Gamma / 2}
        \label{DephasedG} \\
    T_{\varphi}(\ve)
        & = &
            \frac{ \Gamma_{\varphi} \Gamma_{0} }{ (\ve-\ve_d)^2 + (\Gamma/2)^2}
         \label{TPhis} \\
    \alpha(\ve) & = &
            \frac{1}{2} \left[
                \arg \tilde{G}^R(\ve) -
            \arctan \left( \frac{\Gamma_0 - \Gamma_{\varphi}}{2 (\ve-\ve_d)} \right)
            \right]
        \label{AlphaDefn}
\end{eqnarray}
Here, $\Gamma = \Gamma_0 + \Gamma_{\varphi}$ represents the total
width of the level, $\tilde{G}^R(\ve)$ is the retarded Green
function of the level in the presence of dephasing, and
$T_{\varphi}$ parameterizes the strength of the dephasing.
Transmission into the voltage probe from the physical leads will
be described by an $M \times 2$ sub-matrix of $s_{big}$ which we
denote $t_\varphi$. Using a polar decomposition \cite{RMTReview},
it may in general be written as:
\begin{equation}
    \left[t_{\varphi}\right]_{m j} =
      \sum_{k=1,2}   V_{m k}(\ve) \sqrt{T_{\varphi,k}(\ve)} U^T_{k j}(\ve)
\end{equation}
Here, $U$ and $V$ are unitary matrices which parameterize the
preferred modes in the leads, and $T_{\varphi_,k}$ are the two
transmission eigenvalues characterizing the strength of
transmission into the voltage probe.  These transmission
eigenvalues are uniquely specified by $\tilde{s}_0(\ve)$:
\begin{eqnarray}
    T_{\varphi,k}(\ve)  & = &  \delta_{k 1} T_{\varphi}(\ve)
\end{eqnarray}

A general result of voltage probe models is that the average
current $\langle I \rangle$ and current noise $S_I$ are
independent of the matrices $U$ and $V$ appearing in the polar
decomposition \cite{Buttiker,deJong}.  This is convenient, because
in general, these matrices are not uniquely determined by the form
of the coherent scattering matrix $\tilde{s}_0(\ve)$.  However, in
the present problem we are also interested back-action charge
noise $S_Q$, which in general {\it does} depend on these matrices.
For the dephased resonant-level, the matrix $U$ is completely
specified by $\tilde{s}_0(\ve)$:
\begin{equation}
    U(\ve)  =
        e^{i \alpha(\ve) / 2}
        \left( \begin{array}{cc}
             e^{i \alpha(\ve) / 2} \cos \theta &   - e^{-i \alpha(\ve) / 2} \sin \theta \\
             e^{i \alpha(\ve) / 2} \sin \theta &    e^{-i \alpha(\ve) / 2} \cos \theta  \\
    \end{array} \right)
\end{equation}
where the angle $\theta = \tan^{-1} \sqrt{\Gamma_L / \Gamma_R}$
parameterizes the asymmetry in the coupling to the leads.

The matrix $V$ remains unknown; to specify it, we make the
additional assumption that the voltage probe is coupled to the
resonant level via a tunneling Hamiltonian with energy-independent
tunnel matrix elements.  This yields:
\begin{equation}
    V_{m 1} = -i \sqrt{\frac{\Gamma_m}{\Gamma_{\varphi}} } e^{-i
    \alpha(\ve)} e^{i \arg \tilde{G}^R(\ve)},
    \label{VSimple}
\end{equation}
with $\Gamma_{\varphi} = \sum_{m=1}^M \Gamma_m$.  Again, the
ambiguity in $V(\ve)$ has no affect on the average current through
the system or on the current noise.  It will however be important
in determining the charge noise of the system; the choice given in
Eq. (\ref{VSimple}) represents a best case scenario, in that it
minimizes the charge noise.

Finally, we must specify the distribution function in the
reservoir attached to the third lead.  We will contrast three
different choices which all yield a vanishing average current in
the third lead, and which correspond to different physical
mechanisms of dephasing \cite{Buttiker,deJong}. The first
corresponds to a physically-realizable situation where the
reservoir attached to the third lead has a well defined chemical
potential; it is chosen to yield a vanishing average current into
the probe.  We term this the ``pure escape" voltage probe model.
The second model is similar, except one now also enforces the
vanishing of the probe current at each instant of time by allowing
the voltage associated with the third-lead to fluctuate
\cite{BeenakkerButtiker}. This model is usually taken to give a
good description of inelastic effects, and is known as the
inelastic voltage probe model. Finally, in the third model one
also enforces current conservation as a function of energy.  This
is achieved by assigning a non-equilibrium distribution function
to the reservoir \cite{deJong}.  This model is thought to
well-describe quasi-elastic dephasing effects, and is known as the
dephasing voltage probe model.


In what follows, we calculate the average current (and hence the
gain $\lambda$), current noise $S_I$ and charge noise $S_Q$ from
the scattering matrix $s_{big}$ describing the dephased resonant
level detector.  The standard relations between these quantities
and the scattering matrix are given, e.g., in Ref.
\onlinecite{Me}. We will study the effects of dephasing by keeping
the total width of the level $\Gamma$ constant, and varying its
incoherent fraction $\Gamma_{\varphi} / \Gamma$.  This is
equivalent to asking how the noise and detector properties of a
given Lorentzian conductance resonance of fixed width $\Gamma$
depends on the {\it degree} to which it is coherent.  Of course,
simply increasing dephasing while keeping the coupling to the
leads fixed (e.g. by increasing temperature) would also cause the
overall width $\Gamma$ to increase.

\begin{figure}
\center{\includegraphics[width=8.5 cm]{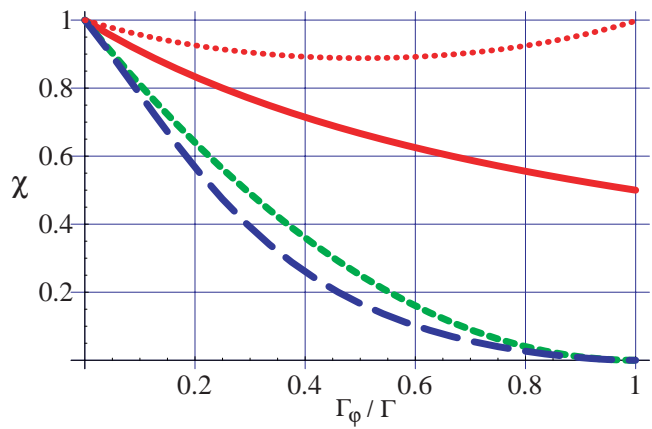}} \vspace{-0.3
cm} \caption{\label{QLPlot1} Quantum efficiency ratio $\chi$
versus $\Gamma_{\varphi}/\Gamma$ for the resonant level model, in
the limit $e V / \Gamma \ra 0$, and with symmetric couplings
$\Gamma_L = \Gamma_R$.  The solid red curve corresponds to the
``pure escape" voltage probe model, the short-dashed green curve
to the inelastic voltage probe model, and the long-dashed blue
curve to the dephasing voltage probe model; each has $\ve_d - \mu
= \Gamma/2$.  In addition, the dotted red curve corresponds to the
``pure escape" model at $\ve_d - \mu = 10 \Gamma$. For strong
dephasing, only the ``pure escape" model is able to remain near
quantum-limited.}
\end{figure}
\begin{figure}
\center{\includegraphics[width=8.5 cm]{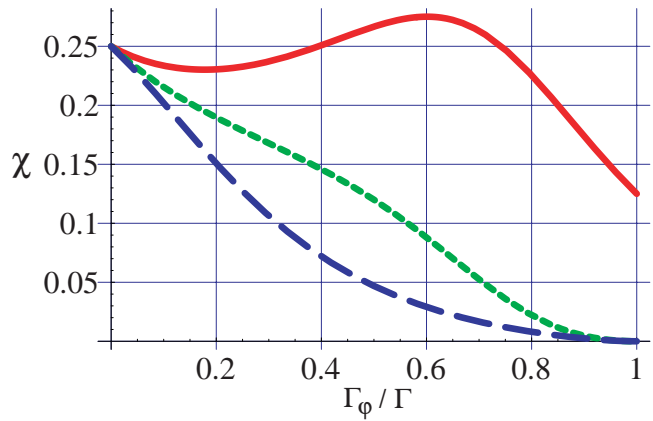}} \vspace{-0.3
cm} \caption{\label{QLPlot2} Quantum efficiency ratio $\chi$
versus $\Gamma_{\varphi}/\Gamma$ for the resonant level model, in
the limit $e V / \Gamma \ra 0$, and with $\Gamma_L \gg \Gamma_R$.
We have chosen $\ve_d = \Gamma/ (2 \sqrt{3})$ to maximize gain.
The different curves correspond to different voltage probe models,
as labelled in Fig. 1.  Note the marked non-monotonic behaviour of
$\chi$ in the ``pure escape" voltage probe model. }
\end{figure}

\subsection{Results from the ``pure escape" voltage probe model}

For simplicity, we focus throughout this subsection on the zero
voltage limit. In general, the average current has both a coherent
contribution (involving only the scattering matrix $\tilde{s}_0$)
and an incoherent contribution, which involves transmission into
the voltage probe lead \cite{Buttiker}.  These combine to yield
the simple result \cite{ButtikerVProbe}:
\begin{eqnarray}
    \langle I \rangle & = &
        \frac{e^2 V}{h} \frac{2 \Gamma_L \Gamma_R}{\Gamma_L + \Gamma_R}
         \left[
        - \textrm{Im } \tilde{G}^R(\ve = \mu) \right] \nonumber \\
    & = &
        \frac{e^2 V}{h}
          \left(\frac{\Gamma_0}{\Gamma} \right)
          \sin^2 2 \theta
             \frac{\Gamma^2 / 4}{(\mu-\ve_d)^2 + \Gamma^2/4}
    \label{PhysVPI}
\end{eqnarray}
The conductance continues to have a Lorentzian form even in the
presence of dephasing, though its overall weight is suppressed by
a factor $(\Gamma_0 / \Gamma)$; the gain $\lambda$ will be
suppressed by the same factor.  Note that this suppression is
indistinguishable from simply enhancing the asymmetry between the
couplings to the leads.

The effect of dephasing on the current noise is more pronounced.
In the present model, even though the {\it average} current into
the voltage probe vanishes, there may nonetheless exist {\it
fluctuating} currents into and out of the voltage probe.  The
result is that measuring different linear combinations of the
current in the left and right lead will yield different values for
the current noise, though they all yield the same average current.
Choosing the measured current to be the linear combination:
\begin{equation}
    I_{meas} = (\sin^2 \alpha) I_L + (\cos^2 \alpha) I_R,
    \label{PhysVPIMeas}
\end{equation}
and writing the current noise in terms of the Fano factor $f$:
\begin{equation}
    S_I = 2 e f \langle I \rangle
\end{equation}
one finds that in zero dephasing case, $f$ is independent of
$\alpha$ and is given by the coherent reflection probability $(1 -
T_0(\ve))$ (c.f. Eq. (\ref{T0Eqn})), whereas in the strong
dephasing limit ($\Gamma_0 / \Gamma \ra 0$), it is given by:
\begin{equation}
    f \ra  \sin^4 \alpha + \cos^4 \alpha
    \label{PhysFano}
\end{equation}
This form reflects the suppression of correlations between $I_L$
and $I_R$; $f$ ranges from a minimum of $1/2$ (for a symmetric
combination of $I_L$ and $I_R$) to a maximum of $1$ if one
measures either $I_L$ or $I_R$.

Turning to the charge noise, we find:
\begin{eqnarray}
    S_Q  =
          \frac{e^3 V \hbar}{2 \pi}
            \left(
                \frac{2 \Gamma_L \Gamma_R}{\Gamma_0^2}
            \right)
            \frac{  \left[
            1 - \left(
                \frac{\Gamma_{\varphi}}{\Gamma}
                \right)^2
            \right] \Gamma^2 }
            { \left[ (\mu-\ve_d)^2 + \Gamma^2/4 \right]^2 }
            \label{PhysVPSQ}
\end{eqnarray}
Within the ``pure escape" voltage probe model, {\it $S_Q$
decreases monotonically with increasing dephasing}, regardless of
asymmetry or the position of the level. As was discussed in Ref.
\onlinecite{Me}, the charge noise $S_Q$ can be regarded as a
measure of the total accessible information in a scattering
detector.  Thus, there is no ``conservation of information" as
dephasing is increased in the present model. In the strong
dephasing limit, $S_Q$ is suppressed in the same way as $\langle I
\rangle$ and $S_I$, that is by a factor $\Gamma_0 / \Gamma$. We
again emphasize that this result corresponds to a {\it
physically-realizable} setup, where a third lead is attached to
the level and assigned a well-defined chemical potential. The
effect of a similar dephasor on the charge fluctuations of a
quantum-point contact was studied experimentally by Sprinzak et.
al in Ref. \onlinecite{Sprinzak}; in contrast to the result of Eq.
(\ref{PhysVPSQ}) for the resonant level detector, they found that
the addition of dephasing did not appreciably change the
back-action noise of a quantum point-contact detector.  Of course,
the system studied in Ref. \onlinecite{Sprinzak} is very different
from the one studied here. Nonetheless, our result indicates that,
at the very least, the insensitivity of charge fluctuations to
dephasing seen in this experiment is not generic to all mesoscopic
conductors.

Finally, turning to the measurement efficiency ratio $\chi$, we
note that since each of $\lambda, S_I$ and $S_Q$ are suppressed as
$\Gamma_0 / \Gamma$, turning on dephasing {\it does not lead to a
parametric suppression of $\chi$} (see Figs. 1 and 2).  In the
strong dephasing limit, we find:
\begin{equation}
    \chi \ra \frac{ 1}{2 f} \frac{(\mu-\ve_d)^2}{(\mu-\ve_d)^2 + \Gamma^2/4},
\end{equation}
where the Fano factor $f$ is given in Eq. (\ref{PhysFano}). In the
strong dephasing limit there is wasted phase information due to
the strong asymmetry between the coupling to the physical leads
and to the voltage probe lead.  Nonetheless, the effects of this
wasted phase information can be minimized by working far from
resonance.  Thus, for dephasing due to true escape (i.e. due to a
simple third lead), {\it one can approach the quantum limit even
in the strongly incoherent limit}.  In this limit, all transport
involves entering the voltage probe, and then leaving it
incoherently.  Nonetheless, in the small voltage limit we
consider, there is no wasted amplitude information in the presence
of dephasing. Though for energies in the interval $[\mu_2,\mu_3]$
there is a current $I_{3,in}$ flowing into the voltage probe lead,
one cannot learn anything new by measuring it, as an {\it
identical} current exits the voltage probe in the energy interval
$[\mu_3,\mu_1]$ and contributes directly to the measured current
flowing between the left and right contacts.  It is this lack of
wasted amplitude information that allows one to reach the quantum
limit even for strong dephasing.

\subsection{Results from the inelastic voltage probe
model}

As discussed earlier, the inelastic voltage probe model is
identical to the ``pure escape" model of the last subsection,
except that we now also require that the current into the voltage
probe vanishes at each instant of time.  This is achieved on a
semi-classical level by assigning the voltage probe a
fluctuating-in-time voltage chosen to exactly enforce this
additional constraint\cite{BeenakkerButtiker}.  The average
current is independent of this additional step, and is identical
to that in the previous section, Eq. (\ref{PhysVPI}). For the
current noise, as we now have $I_L$ = $I_R$ at all times, $S_I$
becomes independent of the particular choice of measured current.
In essence, the effect of the fluctuating voltage in the probe is
to simply choose a particular value of $\alpha$ in Eq.
(\ref{PhysVPIMeas}) \cite{BeenakkerButtiker}. In the strong
dephasing limit, the Fano factor is given by:
\begin{equation}
    f \ra \frac{ \Gamma_L^2 + \Gamma_R^2}{ (\Gamma_L + \Gamma_R)^2
    }
    \label{IncFano}
\end{equation}
Not surprisingly, this is the classical Fano factor corresponding
to two Poisson processes in series.  One also finds this Fano
factor in the large voltage regime of the coherent resonant level
model, where the model may be treated using classical rate
equations \cite{DaviesRLM, AverinOldRLM}.  Note that depending on
the position of $\ve_d$ and the ratio $\Gamma_L / \Gamma_R$, $f$
can either increase or decrease with increasing dephasing.

Finally, we turn to the calculation of the charge noise.  The
effects of the fluctuating voltage on $S_Q$ are far more extreme
than on $S_I$ \cite{Thanks}, as it leads to a new, classical
source for charge fluctuations.  Note that in general we have:
\begin{equation}
    \langle Q \rangle = e \int d \ve \rho(\ve)
        \left(
            \frac{\Gamma_L f_L(\ve) + \Gamma_R f_R(\ve) + \Gamma_\varphi
                f_3(\ve)}{\Gamma}
        \right)
\end{equation}
where $\rho(\ve) = \left(- \frac{1}{\pi} \textrm{Im}
\tilde{G}^R(\ve) \right)$. Fluctuations of the voltage in the
probe lead will cause the probe distribution function $f_3(\ve)$
to fluctuate, and will in turn cause $\langle Q \rangle$ to
fluctuate. Letting $\Delta Q(t)$ denote the fluctuating part of
the charge $Q$, one has \cite{Seelig}:
\begin{equation}
    \Delta Q(t) =
        \left[ \Delta Q(t) \right]_{bare} +
            e \frac{\Gamma_{\varphi}}{\Gamma}
            \int d \ve \rho(\ve) \left(
                 \Delta f_3(\ve,t)
                \right)
    \label{DeltaQ}
\end{equation}
The first term arises from fluctuations in the total current
incident on the scattering region; it is the only contribution
present in the absence of a fluctuating potential, and is
identical to what is found in the ``pure escape" model. The second
term describes fluctuations of $\langle Q \rangle$ arising from
voltage fluctuations in the voltage probe. These voltage
fluctuations are in turn completely determined by the requirement
that the current into the voltage probe vanish at all times
\cite{BeenakkerButtiker}.  Note that as a result, the two terms in
Eq. (\ref{DeltaQ}) will be correlated.

Including these effects, one finds that the charge fluctuations
are given by:
\begin{equation}
    S_Q = \left[S_Q \right]_{bare} + \left[S_Q \right]_{class}
    \label{TwoSQ}
\end{equation}
Here, $[S_Q]_{bare}$ describes the intrinsic charge fluctuations
(i.e. from the first term in Eq. (\ref{DeltaQ})); its value is
given by Eq. (\ref{PhysVPSQ}).  The term $[S_Q]_{class}$ arises
from fluctuations of $\Delta f_3$ (i.e. second term in Eq.
(\ref{DeltaQ})), and includes the effect of their correlation with
$[\Delta Q]_{bare}$.  It is given by:
\begin{equation}
    \left[S_Q \right]_{class} =
        4 e \hbar^2 \frac{\Gamma_{\varphi}} {(\Gamma_0)^2 \Gamma }
        \left(1 + \frac{\Gamma_0 \Gamma_{\varphi}/2} {\omega^2 +
        (\Gamma/2)^2} \right) | \langle I \rangle |
        \label{ExtraSQ}
\end{equation}
where $\langle I \rangle$ is given in Eq. (\ref{PhysVPI}), and
where we have again taken the zero voltage limit.  In the strong
dephasing limit this contribution dominates, and we find:
\begin{equation}
    S_Q \ra
         4 e \hbar^2 \frac{ | \langle I \rangle | } {(\Gamma_0)^2  }
    \label{VPSQLargeDeph}
\end{equation}
Unlike Eq. (\ref{PhysVPSQ}) for the intrinsic charge fluctuations,
which are suppressed with increased dephasing, Eq. (\ref{ExtraSQ})
describing the induced fluctuations of $Q$ diverges in the strong
dephasing limit as $\Gamma / \Gamma_0$.  As a result, the
measurement efficiency $\chi$ tends to zero as $(\Gamma_0 /
\Gamma)^2$, and one is far from the quantum limit for strong
dephasing.  This result is not surprising. The fluctuating voltage
in the probe lead represents a classical uncertainty in the system
stemming from the source of dephasing; this uncertainty in turn
leads to extraneous noise in $Q$, leading departure from the
quantum limit. On a heuristic level, there is unused information
residing in the voltage probe degrees of freedom responsible for
generating the fluctuating potential.  Of course, if this
classical noise is ad-hoc suppressed, one is back at the ``pure
escape" model of the previous section, which can indeed reach the
quantum limit even for strong dephasing.

\subsection{Results from the dephasing voltage probe model}

In the dephasing voltage probe model, the distribution function in
the voltage probe reservoir is chosen so that the current flowing
into the voltage probe vanishes at each energy \cite{Buttiker}.
One obtains a non-equilibrium distribution function $f_3(\ve)$
defined by:
\begin{equation}
    f_3(\ve) = \bar{f}(\ve) \equiv \frac{ \Gamma_L f_L(\ve) + \Gamma_R f_R(\ve) }
    {\Gamma_L + \Gamma_R}
    \label{fbarDefn}
\end{equation}
Similar to the inelastic voltage probe, we also enforce a
vanishing current into the voltage probe at each instant of time
by having $f_3(\ve)$ fluctuate in time \cite{deJong}.  This model
is usually thought to better mimic pure dephasing effects than the
inelastic voltage probe, as the coupling to the voltage probe does
not lead to a redistribution of energy.

For the average current in the small voltage regime, we again
obtain Eq. (\ref{PhysVPI}), as in the ``pure escape" model. In the
finite voltage regime we obtain the simple result:
\begin{equation}
    \langle I \rangle =
        \frac{e}{h} \frac{2 \Gamma_L \Gamma_R}{\Gamma_L + \Gamma_R}
        \int d\ve \left( f_1(\ve)  - f_2 (\ve) \right) \left(
        - \textrm{Im } \tilde{G}^R(\ve) \right),
    \label{DephVPI}
\end{equation}
where again $\tilde{G}^R(\ve)$ is the Green function of the level
in the presence of dephasing (c.f. Eq. (\ref{DephasedG})).  Eq.
(\ref{DephVPI}) is identical to the formally exact expression
derived by Meir and Wingreen \cite{Meir} for the current through a
single level having {\it arbitrary} on-site interactions.  Thus,
we can think of the dephased Green function $\tilde{G}^R$ in the
voltage probe approach as mimicking the effects of dephasing due
to interactions.  The simplicity of both the Meir-Wingreen and
voltage-probe approaches (i.e. $\langle I \rangle \propto G^R$)
may be simply extended to the multi-level case if the asymmetry in
the tunnel couplings is the same for each level.

Turning to the noise, the main effect of the non-equilibrium
distribution function $f_3$ will be to {\it increase} the noise
over the previous two models. On the level of the calculation,
these new contributions arise from diagonal elements of the
current $I$ and charge $Q$ operators, considered in the basis of
scattering states.  More physically, this additional noise is due
to the fact that the detector plus voltage probe system is no
longer described by a single pure state. Rather, the distribution
function $f_3$ corresponds to a statistical ensemble of states,
with each state of the ensemble yielding a (possibly) different
quantum expectation of $Q$ and $I$.  The extra noise produced by
this ``classical" uncertainty corresponds to unused information
(e.g. if the non-purity of the detector density matrix results
from entanglement with a reservoir, the missing information
resides in the reservoir degrees of freedom), and thus we
anticipate a departure from the quantum limit {\it even} if we
neglect the extraneous charge noise arising from the fluctuating
voltage probe potential.

We find that the modification of the current noise is rather
minimal-- in the strong dephasing limit, one again obtains the
classical Fano factor of Eq. (\ref{IncFano}).  The effect on the
charge noise is more pronounced.  Similar to the inelastic voltage
probe, there will be two contributions to $S_Q$, one stemming from
fluctuations of the total incident current on the level, the other
from fluctuations of the distribution function $f_3(\ve)$.
Separating the two contributions to $S_Q$ as in Eq. (\ref{TwoSQ}),
we find for {\it arbitrary} voltages:
\begin{equation}
    [S_Q]_{bare} = \frac{4 e^2 \hbar}{\pi}
        \int d \ve
            \bar{f}(\ve) \left(1 - \bar{f}(\ve) \right)
        \left(
            -\textrm{ Im } \widetilde{G}^R(\ve)
        \right)^2
    \label{DephVPSQ}
\end{equation}
\begin{eqnarray}
    \left[S_Q \right]_{class} =
    \frac{ 4 e^2 \hbar}{\pi}
        \frac{\Gamma_{\varphi}}{\Gamma \Gamma_0} \int d \ve
            \bar{f}(\ve) \left(1 - \bar{f}(\ve) \right)
            && \nonumber
            \\
            \left(
            -\textrm{ Im } \widetilde{G}^R(\ve)
        \right)
        \left( 1 +
            \frac{2 \Gamma_0 \Gamma_{\varphi}}
                {\ve^2 + (\Gamma/2)^2} \right)
                \label{DephVPSQClass}
\end{eqnarray}
Interestingly, we find that in the dephasing voltage probe model,
the intrinsic charge fluctuations $[S_Q]_{bare}$ (Eq.
(\ref{DephVPSQ})) are independent of dephasing strength; the
additional noise due to the non-equilibrium distribution function
$\bar{f}$ exactly compensates the suppression of the intrinsic
noise found in the ``pure escape" model and inelastic voltage
probe models (c.f. Eq. (\ref{PhysVPSQ})).  Recall that $S_Q =
[S_Q]_{bare}$ in the absence of voltage fluctuations in the
voltage probe lead.

The extrinsic contribution $[S_Q]_{class}$ (Eq.
(\ref{DephVPSQClass})) is similar in form to the corresponding
contribution for the inelastic voltage probe model.  In the strong
dephasing limit this term dominates, and at arbitrary voltage, we
again obtain the divergent result of Eq. (\ref{VPSQLargeDeph}) for
$S_Q$. Similar to the inelastic voltage probe model, the
fluctuations of $Q$ induced by the fluctuating probe voltage will
cause a suppression of $\chi$ as $(\Gamma_0 / \Gamma)^2$ in the
strong dephasing limit. Note that even if one did not use a
fluctuating voltage in the voltage probe (i.e. enforce current
conservation in energy but not in time), and hence had only the
intrinsic contribution to $S_Q$, one would still have a parametric
suppresion of $\chi$ at strong dephasing; this is in contrast to
the inelastic voltage probe model, where it is only
$[S_Q]_{class}$ that prevents reaching the quantum limit at strong
dephasing. Including all terms in $S_Q$, one finds that the
meaurement efficiency for the dephasing voltage probe model {\it
is always less than} that for the inelastic dephasing voltage
probe model (see Figs. 1 and 2). As discussed above, these result
follows from the additional classical uncertainty resulting from
the voltage probe plus detector system being in a mixed state.

\begin{figure}
\center{\includegraphics[width=8.5 cm]{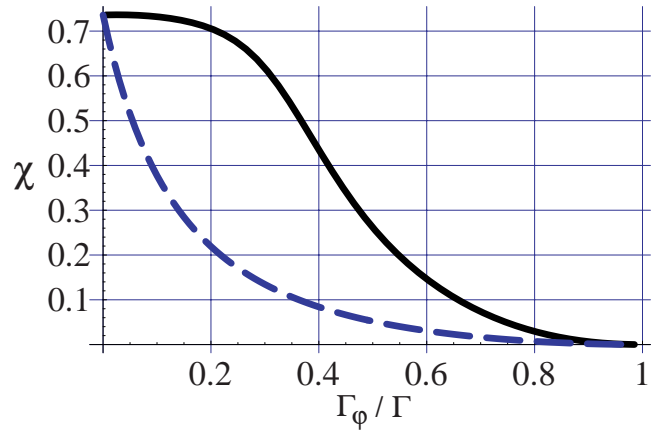}} \vspace{-0.3
cm} \caption{ Quantum efficiency ratio $\chi$ versus
$\Gamma_{\varphi}/\Gamma$ for the resonant level model, in the
large-voltage ``cotunneling" limit:  $\mu_L-\mu_R = 100 \Gamma$,
and $\ve_d = \mu_L + \Gamma$. The long-dashed blue curve
corresponds to the dephasing voltage probe model, while the solid
black curve corresponds to the slow random potential model (with
$\Gamma_{\varphi} \ra \lambda_{\varphi}$). Despite the large
voltage, $\chi$ is still suppressed to zero by dephasing in both
models; note also the large difference between the two models.}
\label{QLPlotLargeV}
\end{figure}

Finally, it is interesting to consider the case of a large voltage
($\mu_L - \mu_R = e V \gg \Gamma$) ``co-tunneling" regime, where
the level is placed slightly above the higher chemical potential
$\mu_L$. Averin \cite{AverinRLM} demonstrated that in the fully
coherent case, one can still come close to the quantum limit (i.e.
$\chi \ra 3/4$) in this regime despite the loss of information
associated with the large voltage and the consequent energy
averaging.  One might expect $\chi$ to be insensitive to dephasing
in this regime, given the large voltage.  This is not the case; as
is shown in Fig. \ref{QLPlotLargeV}, $\chi$ is again suppressed to
zero as $(\Gamma_0 / \Gamma)^2$.

\section{Dephasing From a Fluctuating Potential}

\subsection{General setup}

We now consider an alternate model of dephasing in which the
resonant level detector is subject to a random,
Gaussian-distributed, time-dependent potential.  This model
represents the classical (high-temperature) limit of dephasing
induced by a bath of oscillators (e.g. phonons), and is attractive
as it allows a simple semiclassical interpretation of the
influence of dephasing on noise.  It also allows one to make a
clear distinction between pure dephasing effects and inelastic
scattering.  Surprisingly, we find that for noise properties,
dephasing from a fluctuating random potential is not completely
equivalent to any of the voltage probe models, even if one chooses
a ``slow" potential which does not give rise to inelastic effects;
one finds a reasonable agreement between the models only in the
small voltage regime. The upshot of the analysis is that the noise
properties and detector efficiency of the resonant level model is
sensitive both to the degree of detector coherence {\it and} the
nature of the dephasing source.  Study of the random potential
model also helps elucidate the origin of the two contributions to
the charge noise found in voltage probe models (c.f. Eq.
(\ref{TwoSQ})).

Note that a model somewhat similar to that considered here was
used by Davies et. al \cite{DaviesIncRLM} to study the effect of
dephasing on current noise in a double tunnel-junction structure.
Unlike the present study, they focused on the large voltage regime
$e V \gg \Gamma$, and could not consider the effect of varying the
timescale of the random potential.  The effect of a fluctuating
potential on current noise in a mesoscopic interferometer was also
recently studied by Marquardt et. al. \cite{Marquardt}. The
situation here is quite different, as the scattering has a marked
energy dependence, and we are also interested in the back-action
charge fluctuations.

The time-dependent Hamiltonian for the system is:
\begin{eqnarray}
    H(t) & = & \left[ \ve_d + \eta(t)\right] d^{\dag} d^{\pd}
        +
        \sum_{\alpha=L,R} \int_{-D}^D d \epsilon  \left[ \epsilon \cdot
        c^{\dag}_{\alpha}(\epsilon)
        c^{\pd}_{\alpha}(\epsilon) \right]
        +
        \nonumber \\
        &&
  \sum_{\alpha=L,R} \sqrt{\frac{\Gamma_{\alpha}}{2 \pi}}
     \int d \epsilon \left[
            d^{\dag} c^{\pd}_{\alpha} (\epsilon) +
            c^{\dag}_{\alpha}(\epsilon)
            d^{\pd} \right]
\end{eqnarray}
Assuming that the conduction electron bandwidth $D$ is the largest
scale in the problem, we may solve the Heisenberg equations.
Setting $\hbar = 1$, we find:
\begin{equation}
    d(t) = \int d t' G^R(t, t')
        \sum_{\alpha} \sqrt{ \frac{\Gamma_{\alpha}}{2 \pi}}
        \int d \ve e^{-i \ve t'}
        \tilde{c}^{\pd}_{\alpha}(\ve)
        \label{RndDEqn}
\end{equation}
where
\begin{eqnarray}
    G^R(t,t') & = & -i \theta(t-t') e^{- \Gamma_0 (t-t') / 2}
        e^{-i \ve_d (t-t')} \times \nonumber \\
        && \exp\left( -i \int_{t'}^{t} d \tau \eta(\tau) \right)
        \label{GRTime}
\end{eqnarray}
and where the $\tilde{c}_{\alpha}(\ve)$ operators describe
conduction electrons in the leads in the absence of tunneling.
Expectations of the $\tilde{c}_{\alpha}$ operators obey Wick's
theorem, and are given in terms of the lead distribution functions
in the usual manner.  We will assume throughout that the noise
$\eta(t)$ is stationary and Gaussian with an auto-correlation
function $J(\tau)$:
\begin{equation}
    \langle \eta(t) \eta(t') \rangle = J(t-t') = J(t'-t)
\end{equation}
Note that whereas voltage probe models are essentially
characterized by a single energy scale $\Gamma_{\varphi}$, here
the bandwidth and magnitude of $J(t=0)$ give two distinct scales.

\subsubsection{Average current}

Defining the current in lead $\alpha$ as the time derivative of
the particle number in lead $\alpha$ yields:
\begin{equation}
    I_{\alpha}(t)/e = - \Gamma_\alpha d^{\dagger}(t)
    d(t) + i  \sqrt{\frac{ \Gamma_{\alpha}}{2
    \pi}} \int d \omega \left[
        e^{i \omega t} \tilde{c}^{\dag}_{\alpha}(\omega) d(t) -
        \textrm{ h.c. } \right]
    \label{RndI}
\end{equation}
Taking the expectation of Eq. (\ref{RndI}) for the current in the
right lead,  we find:
\begin{widetext}
\begin{eqnarray}
    \langle I_\alpha(t) \rangle & = & -e \Gamma_\alpha
        \int d \omega \int_{-\infty}^{\infty} d t_0 \Bigg[
            \Gamma_0 \bar{f}(\omega)
            \int_{- \infty}^{\infty}
            d t_b
            G^A(t_b,t;\omega) G^R(t,t_0;\omega) +
             2 f_\alpha(\omega) \textrm{ Im}
         G^R(t,t_0; \omega) \Bigg]
    \label{RndAvgI1}
\end{eqnarray}
\end{widetext}
where $\bar{f}$ is defined in Eq. (\ref{fbarDefn}), and the
additional argument $\omega$ in $G^{R/A}$ indicates that one
should shift $\ve_d \ra \ve_d - \omega$ in Eq. (\ref{GRTime}).
Given the simple form of $G^R(t,t')$, there is an optical theorem
which relates these two contributions for arbitrary $\eta(t)$:
\begin{eqnarray}
    && \int d t_a \int d t_b
            G^A(t_b,t; \omega) G^R(t,t_a;\omega)  =
                \label{Optical1}  \\
     &&       - 2  \int d {t_0}
                    e^{-\Gamma_0 t_0}
                \int d \tau \textrm{ Im } G^R(t-t_0, t-t_0-\tau;\omega)
                \nonumber
\end{eqnarray}
Using this to simplify Eq. (\ref{RndAvgI1}) and then averaging
over the random potential $\eta$ yields:
\begin{eqnarray}
    \langle \langle I_R \rangle \rangle_\eta = &&
    \frac{e}{2 \pi}
    \frac{2 \Gamma_R \Gamma_L}{\Gamma_0}
        \int d \omega ( f_R(\omega) - f_L(\omega) )
    \nonumber \\
    &&
        \int_0^{\infty} d \tau
        \textrm{ Im }\langle G^R(\tau,0; \omega) \rangle_\eta
    \label{RndAvgI2}
\end{eqnarray}
We have used the fact that the $\eta$-averaged value of $G^R$ is
invariant under time-translation.  Eq. (\ref{RndAvgI2}) for
$\langle I \rangle$ is identical to the expression emerging from
the dephasing voltage probe model (c.f. Eq. (\ref{DephVPI})), with
the $\eta$-averaged Green function playing the role of the
``dephased" Green function $\tilde{G}^R$ in the latter model.  On
a heuristic level, Eq. (\ref{RndAvgI2}) indicates that tunneling
processes with different dwell times $\tau$ on the level
contribute to $\langle I \rangle$; the random phases picked up
during these events will cause a suppression of the current.

Averaging the $\eta$ dependent parts of $G^R$ yields:
\begin{eqnarray}
    \langle G^R(\tau) \rangle_{\eta} = &&
                e^{-i \ve_d \tau} e^{- \Gamma_0 \tau / 2}
                \times \label{RndAvgG} \\
    &&
            \exp \left(
            -\frac{1}{\pi } \int_{-\infty}^{\infty} d \omega
            \frac{\sin(\omega t / 2)^2}{\omega^2} J(\omega)
            \right) \nonumber
\end{eqnarray}
Not surprisingly, the factor in $\langle G^R \rangle$ arising from
the averaging has an identical form to what is encountered when
studying the dephasing of a spin coupled to a random potential or
to a bosonic bath.

\subsubsection{Charge noise}

In general, the noise in a quantity $X$ in the fluctuating
potential model will have two distinct sources:
\begin{equation}
    \textrm{Var} \left[X^2\right]  =
        \Big\langle
             \left[ \langle X^2 \rangle - \langle X \rangle^2
             \right]
        \Big\rangle_{\eta} +
        \left(
            \Big \langle \langle X \rangle ^2 \Big \rangle_{\eta}
              - \langle \langle X \rangle \rangle_\eta ^2 \right)
        \label{TwoNoises}
\end{equation}
The first term describes the intrinsic noise for each given
realization of $\eta(t)$, whereas the second describes the
classical fluctuation of the average value of $X$ from realization
to realization.  In what follows, we will focus on the first, more
intrinsic effect; this corresponds to an experiment where the
noise (i.e. variance) is calculated for each realization of
$\eta(t)$, and is only then averaged over different realizations.
Note that the neglected classical contribution to the noise (i.e.
the second term in Eq. (\ref{TwoNoises})) is positive definite;
including it will only push the system further from the quantum
limit.

Noting that here, $Q(t) \equiv d^{\dag}(t) d(t)$, we find using
Eq. (\ref{RndDEqn}):
\begin{eqnarray}
    S_Q & \equiv &2
            \int_{-\infty}^{\infty} dt \Big\langle \left[
            \langle Q(t) Q(0) \rangle -
            \langle Q(t) \rangle \langle Q(0)\rangle
            \right] \Big\rangle_{\eta} \nonumber \\
        & = & \frac{2 e^2}{ \pi^2}
            \int_{-\infty}^{\infty} dt
            \int_{-\infty}^{\infty} d \bar{\omega}
            \int_{-\infty}^{\infty} d (\Delta \omega) e^{i \Delta \omega t}
            \int_0^{\infty} d \tau_1   \int_0^{\infty} d \tau_2
            \nonumber \\
        &&
            \bar{f}(\bar{\omega} + \Delta \omega/2)
            \left(1 - \bar{f}(\bar{\omega} - \Delta \omega/2)
            \right)
            e^{-i \Delta \omega (\tau_1 - \tau_2) / 2}
            \nonumber \\
        &&
            \Big\langle
            \textrm{ Im } G^R(t,t-\tau_1;\bar{\omega})
            \textrm{ Im } G^R(0,-\tau_2;\bar{\omega})
        \Big\rangle_{\eta}
        \label{FullRndSQ}
\end{eqnarray}
We have used the optical theorem of Eq. (\ref{Optical1}) to
express $S_Q$ in terms of a product of two (as opposed to four)
Green functions.  Similar to Eq. (\ref{RndAvgI2}) for the current,
Eq. (\ref{FullRndSQ}) for $S_Q$ may be given a simple heuristic
interpretation. The first factor of $G^R$ in Eq. (\ref{FullRndSQ})
yields the amplitude of an event where an electron leaves the
level at time $t$ after having spent a time $\tau_1$ on the level,
while the second describes a process where an electron leaves the
level at time $0$ after a dwell time $\tau_2$. Eq.
(\ref{FullRndSQ}) thus expresses the charge noise as a sum over
pairs of tunneling events occurring at different times. Unlike the
average current (c.f. Eq. (\ref{RndAvgI1})), the charge noise is
sensitive to interference between tunnel events which have
different exit times from the level.

It is useful to write the Green function factor in Eq.
(\ref{FullRndSQ}) as:
\begin{eqnarray}
    \Big\langle
        \textrm{ Im } G^R(t,t-\tau_1;\bar{\omega})
        \textrm{ Im } G^R(0,-\tau_2;\bar{\omega})
    \Big\rangle_{\eta} = &&
        \nonumber \\
        \textrm{Re }
        \frac{
              W^{D}(t;\tau_1,\tau_2) - W^I(t;\tau_1,\tau_2)
        }{2} &&
\end{eqnarray}
where
\begin{subequations}
\label{WDefs}
\begin{eqnarray}
    W^D(t;\tau_1,\tau_2) & = &
                   \Big \langle G^R(t,t-\tau_1) G^A(0,-\tau_2) \Big\rangle
               \\
    W^I(t;\tau_1,\tau_2) & = &
                   \Big\langle G^R(t,t-\tau_1) G^R(0,-\tau_2) \Big\rangle
\end{eqnarray}
\end{subequations}
Similar to standard disorder-averaged calculations, we have both a
``diffuson" ($W^D$) and an ``interference" ($W^I$) contribution.
Averaging over the random potential yields:
\begin{eqnarray}
      W^{D/C}(t;\tau_1,\tau_2)
        =
         e^{-i (\ve_d-\bar{\omega}) ( \tau_1 \mp  \tau_2)}
        e^{- \Gamma_0 (\tau_1 + \tau_2)/2} &&
          \nonumber \\
        \exp \left[ -\frac{1}{2 }
            \int_{-\infty}^{\infty} d t_1 \int_{-\infty}^{\infty} d t_2
            \chi_{\mp}(t_1) \chi_{\mp}(t_2)
            J(t_1-t_2) \right] &&
    \label{AveragedK}
\end{eqnarray}
where
\begin{subequations}
\begin{eqnarray}
        \chi_{\pm}(t) & = &
            \chi_{[t-\tau_1,t]}(t) \pm \chi_{[-\tau_2,0]}(t) \\
        \chi_{[t_i,t_f]}(t) & = & \theta(t-t_i) - \theta(t-t_f)
\end{eqnarray}
\end{subequations}
In what follows, it is useful to make the shift $t \ra t +
(\tau_1-\tau_2)/2$ in Eq. (\ref{FullRndSQ}).  Simplifying, and
using the fact that the $\eta$-averaged Green function is
time-translation invariant, we then have:
\begin{subequations}
    \label{FinalWs}
\begin{eqnarray}
      W^{D}(t;\tau_1,\tau_2)
        & = &
            C(t;\tau_1,\tau_2)
               \left[
                   \langle G^R(\tau_1) \rangle \langle G^A(\tau_2) \rangle
               \right] \\
      W^I(t;\tau_1,\tau_2)
        & = &
           \frac{
                \left[
                    \langle G^R(\tau_1) \rangle \langle G^R(\tau_2) \rangle
                \right] }
            {C(t;\tau_1,\tau_2)}
        \label{AvgK2}
\end{eqnarray}
\end{subequations}
with
\begin{eqnarray}
    C(t;\tau_1,\tau_2) & = &
                 \exp \Big[
                 \int_{-\infty}^{\infty} d t_1 \int_{-\infty}^{\infty} d t_2
                 \nonumber \\
        &&
                 \chi_{[t-\frac{\tau_1+\tau_2}{2},t+\frac{\tau_1-\tau_2}{2}]}(t_1)
                 \chi_{[-\tau_2,0]}(t_2)
                 J(t_1-t_2) \Big] \nonumber \\
        & = &
            \exp \Bigg[
                \frac{2}{\pi }
                 \int_{-\infty}^{\infty} d \omega J(\omega)
                \nonumber \\
            &&
                \times
                 \frac{
                    \sin(\omega \tau_1 /2) \sin(\omega
                    \tau_2/2)}{\omega^2}
                 \cos
                    \omega t
            \Bigg]
            \label{CFactor}
\end{eqnarray}

Eq. (\ref{AveragedK}) indicates that the general effect of the
random potential is to suppress the contribution of each pair of
tunnel events to $S_Q$.  The correlation between the random phases
acquired by each of the two events plays a central role, and is
described by the factor $C(t;\tau_1,\tau_2)$, where
$\tau_1,\tau_2$ are the dwell times of the two events, and $t$ is
the difference in exit times.  Eq. (\ref{CFactor}) gives a simple
expression for $C(t;\tau_1,\tau_2)$ in terms of the spectral
density of the random potential.  The existence of phase
correlations allows pairs of tunnel events to interfere
constructively when contributing to the diffuson contribution;
thus, as can be seen in Eqs. (\ref{FinalWs}), phase correlations
tend to enhance the diffuson contribution relative to the
interference contribution.

In addition, the phase correlation factor $C(t;\tau_1,\tau_2)$ is
the only $t$ dependent factor remaining in the expression for
$S_Q$ after $\eta$ averaging. Thus, it will completely determine
the contribution of inelastic processes to the charge noise $S_Q$
(i.e. terms with $\Delta \omega \neq 0$ in Eq. (\ref{FullRndSQ})).
Such processes correspond to the absorption or emission of energy
by the random potential. Note the formal similarity between the
correlation kernel $C(t;\tau_1,\tau_2)$ in Eq. (\ref{FullRndSQ}),
and the kernel appearing in the $P(E)$ theory describing the
effect of environmental noise on electron tunneling
\cite{PEReference}. Here, not surprisingly, the probability of an
inelastic transition depends on the dwell times $\tau_1$, $\tau_2$
of the two tunnel events.

We can now straightforwardly identify the ``pure dephasing" (i.e.
elastic) contribution to $S_Q$ for an arbitrary $J(\omega)$ by
keeping only the $t$-independent part of $C(t;\tau_1,\tau_2)$ when
evaluating Eq. (\ref{FullRndSQ}).  This is equivalent to replacing
$C(t;\tau_1,\tau_2)$ by $C(t \ra \infty;\tau_1,\tau_2)$ in Eq.
(\ref{FullRndSQ}).  We thus define:
\begin{eqnarray}
    S_Q \big|_{elast} & = & \frac{2 e^2}{\pi} \int_{-\infty}^{\infty}
         d \omega \bar{f}(\omega) (1- \bar{f}(\omega))
            \int d \tau_1 \int d \tau_2
            \label{RndSQElast} \\
     &&
            \textrm{Re }
            \left[ W^D(t \ra \infty;\tau_1,\tau_2) - W^I(t \ra \infty;\tau_1,\tau_2)
            \right] \nonumber
\end{eqnarray}
If, in addition, we have the reasonable result that there are no
phase correlations in the long time-separation limit (i.e.
$C(t;\tau_1,\tau_2) \ra 1$ as $t \ra \infty$), the elastic
contribution becomes:
\begin{equation}
    S_Q \big|_{elast}  =  \frac{4 e^2}{\pi} \int_{-\infty}^{\infty}
         d \omega \bar{f}(\omega) (1- \bar{f}(\omega))
         \Big \langle
                \textrm{Im } G^R(\omega)
            \Big \rangle ^2
        \label{RndSQNoC}
\end{equation}
This expression for the elastic contribution to $S_Q$ is {\it
identical} to what is obtained for the intrinsic charge
fluctuations $[S_Q]_{bare}$ in the dephasing voltage probe model
(c.f. Eq. (\ref{DephVPSQ})), if we associate the $\eta$-averaged
Green function with the dephased Green function $\tilde{G}^R$ in
the latter model.  Recall that the in the voltage probe model, the
intrinsic contribution $[S_Q]_{bare}$ arises from fluctuations of
the total current incident on the level, and is independent of
fluctuations of the probe voltage.
We thus have that if there are no long-time phase correlations,
the purely elastic effect of a random potential is captured (in
form) by the intrinsic charge fluctuations in the dephasing
voltage probe model. Note that in Eq. (\ref{RndSQNoC}) both the
interference and diffuson terms contribute equally.  The lack of
any phase correlations implies that there is no relative
enhancement of $W^D$ over $W^I$.

Finally, we may define a purely inelastic contribution to $S_Q$
for an arbitrary $J(\omega)$ by simply subtracting off the elastic
contribution defined in Eq. (\ref{RndSQElast}) from the full
expression for $S_Q$. We find:
\begin{widetext}
\begin{eqnarray}
    S_Q |_{inelast}  & = & \frac{2 e^2}{\pi}
            \int_{-\infty}^{\infty} d \bar{\omega}
            \int_{-\infty}^{\infty} d (\Delta \omega)
            \int_0^{\infty} d \tau_1   \int_0^{\infty} d \tau_2
            \bar{f}(\bar{\omega} + \Delta \omega/2)
            \left(1 - \bar{f}(\bar{\omega} - \Delta \omega/2)
            \right)
                    \label{RndSQInelast} \\
        &&
            \textrm{ Re} \Bigg[
                P_D(\Delta \omega;\tau_1,\tau_2)
            \Big\langle G^R(\tau_1;\bar{\omega})\Big\rangle_{\eta}
            \Big\langle G^A(\tau_2;\bar{\omega})
            \Big\rangle_{\eta}
            - P_I(\Delta \omega;\tau_1,\tau_2)
            \Big\langle G^R(\tau_1;\bar{\omega})\Big\rangle_{\eta}
            \Big\langle G^R(\tau_2;\bar{\omega}) \Big\rangle_{\eta}
            \Bigg]
            \nonumber
\end{eqnarray}
\end{widetext}
where we have defined the real-valued functions:
\begin{eqnarray}
    && P_{D/I}(\Delta \omega; \tau_1, \tau_2)
        =   \label{PEDefn}\\
    &&        \int dt \frac{e^{i \Delta \omega t}}{2 \pi}
            \Big[ C(t;\tau_1,\tau_2)^{\pm 1} -
            C(t\ra\infty;\tau_1,\tau_2)^{\pm1}
            \Big] \nonumber
\end{eqnarray}
Though they are not necessarily positive definite, the functions
$P_{D/I}(\Delta \omega; \tau_1, \tau_2)$ may be interpreted as the
quasi-probability of obtaining an inelastic contribution of size
$\Delta \omega$ from (respectively) the diffuson or interference
contribution.

\subsubsection{Current noise}

We again focus on the ``intrinsic" fluctuations (e.g. the first
term in Eq. (\ref{TwoNoises})) and ignore the additional
contribution to the current noise arising from variations of
$\langle I \rangle$ in different realizations of the random
potential $\eta$.  Similar to the case of the charge noise, the
current noise $S_I$ may be expressed in terms of products of $G^R$
and $G^A$ at different times.  Also, a direct calculation shows
that current conservation holds for the fluctuations:  the current
noise is independent of the lead in which it is calculated.
Writing the current noise as:
\begin{eqnarray}
    S_I & = & \frac{2 e^2 \Gamma_L \Gamma_R}{2 \pi^2}
            \sum_{\alpha,\beta = L,R}
            \int_{-\infty}^{\infty} d \bar{\omega}
            \int_{-\infty}^{\infty} d (\Delta \omega)
            \int_0^{\infty} d \tau_1   \int_0^{\infty} d \tau_2
            \nonumber \\
            &&
            e^{i \Delta \omega t}
            \left(
                f_{\alpha}(\bar{\omega} + \Delta \omega/2)
            \right)
            \left(1-
                f_{\beta}(\bar{\omega} - \Delta \omega/2)
            \right)  \nonumber \\
            &&
            \times e^{-i \Delta \omega (\tau_1 - \tau_2) / 2}
                        \textrm{ Re}\left[
                \delta_{\alpha \beta} S_{I}^{d} +
                (1-\delta_{\alpha \beta}) S_{I}^{od}
            \right]
\end{eqnarray}
we find after averaging:
\begin{eqnarray}
    S_{I}^{d}(t;\tau_1,\tau_2)
        & = & \frac{2 \Gamma_L \Gamma_R }{\Gamma_0^2}
            \left(
                W^{D} -
                W^I
            \right)
            \label{RndSIo} \\
    S_{I}^{od}(t;\tau_1,\tau_2)
        & = & \frac{
                    \Gamma_L^2 +  \Gamma_R^2  }
            {\Gamma_0^2}
                W^{D} +
          \frac{2 \Gamma_L \Gamma_R }{\Gamma_0^2}
                W^I.
    \label{RndSIod}
\end{eqnarray}
Similar to the charge noise, the current noise results from the
interference between pairs of tunnel events, and may be expressed
in terms of the diffuson and interference terms defined in Eqs.
(\ref{WDefs}) (we have suppressed their time arguments above for
clarity). Unlike the charge noise, we see that for the
off-diagonal fluctuations ($S_{I}^{od}$), the diffuson and
interference term enter with different coefficients. Note that in
the zero voltage case, the interference term $W^I$ does not
contribute:
\begin{equation}
      \left[
          \delta_{\alpha \beta} S_{I}^{d} +
          (1-\delta_{\alpha \beta}) S_{I}^{od}
      \right]
    \ra
         W^D(t;\tau_1,\tau_2)
\end{equation}
Elastic and inelastic contributions to $S_I$ may be identified in
the same way as was done for $S_Q$.


\subsection{Results from the slow random potential model}

\begin{figure}
 \center{\includegraphics[width=8.2
cm]{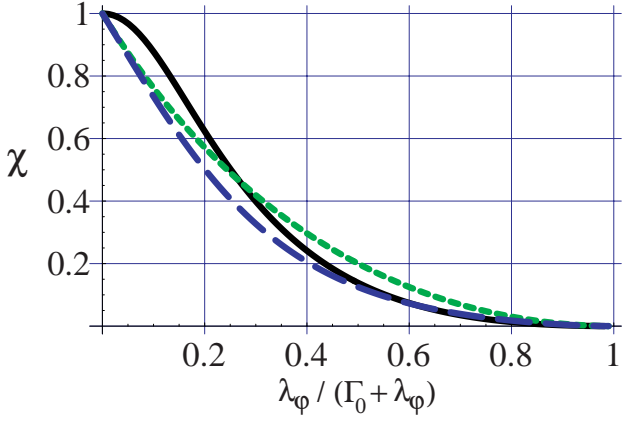}} \vspace{-0.3 cm} \caption{\label{QLPlot4} Solid
black curve: quantum efficiency ratio $\chi$ versus dephasing
strength for the slow random potential model, in the limit $e V
\ra 0$. We have assumed symmetric couplings $\Gamma_L = \Gamma_R$,
and have chosen $\ve_d$ so as to maximize the gain.  For
comparison, the long-dashed blue curve (short-dashed red curve)
corresponds to the dephasing (inelastic) voltage probe model, with
$\lambda_{\varphi} \ra \Gamma_{\varphi}$.}
\end{figure}
\begin{figure}
\label{QLPlot5}\center{\includegraphics[width=8.5
cm]{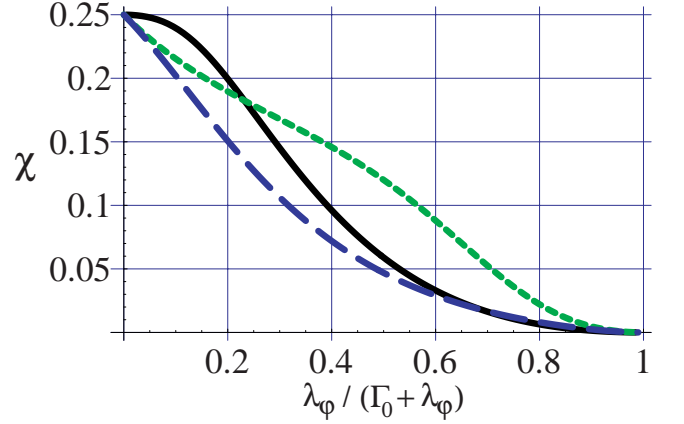}} \vspace{-0.3 cm} \caption{\label{QLPlot5}
Quantum efficiency ratio $\chi$ versus dephasing strength for the
slow random potential, in the limit $e V \ra 0$ and for $\Gamma_L
\gg \Gamma_R$.  We have chosen $\ve_d$ so as to maximize the gain.
Curves labelled as in Fig. 4.}
\end{figure}

Similar to Ref. \onlinecite{Marquardt}, we will consider in what
follows two limiting cases for the spectral density $J(\omega)$.
The first is that of a ``slow" random potential, where the
bandwidth $\Omega$ of $J(\omega)$ is much smaller than the
frequency scales of interest, i.e..
\begin{equation}
    \Omega \ll \Gamma_0, \ve_d-\mu_L, \ve_d-\mu_R,
        |\mu_L - \mu_R|
    \label{OmegaBound}
\end{equation}
This allows us to make the approximation
\begin{equation}
    J(t) \ra J(t = 0)
    \label{SlowApprox}
\end{equation}
Inelastic effects should be minimal in this limit, and thus one
might expect results which are similar to the dephasing voltage
probe model.

For the averaged Green function, the approximation of Eq.
(\ref{SlowApprox}) in Eq. (\ref{RndAvgG}) yields:
\begin{equation}
    \langle G^R(\tau) \rangle =
        e^{-i \ve_d \tau} e^{- \Gamma_0  \tau / 2}
        e^{-(\lambda_\varphi t)^2/2}
    \label{RndAvgGSlow}
\end{equation}
where
\begin{equation}
    \lambda_{\varphi} = \sqrt{J(t \ra 0) }.
\end{equation}
Plugging this into Eq. (\ref{RndAvgI2}) yields a Voit lineshape
(i.e. a convolution of a Gaussian and a Lorentzian):
\begin{eqnarray}
    \langle \langle I \rangle \rangle & = & \frac{e}{2 \pi}
    \frac{2 \Gamma_R \Gamma_L}{\Gamma_0}
        \int d \omega ( f_L(\omega) - f_R(\omega) )
        \nonumber \\
        &&
        \int d \omega'
        \left(
        \frac{e^{-\omega'^2/ (2 \lambda_{\varphi}^2) } }
            {\sqrt{2 \pi \lambda_{\varphi}^2 }} \right)
        \frac{ \Gamma_0/2}{ (\ve_d -
        \omega - \omega')^2 + (\Gamma_0)^2/4} \nonumber \\
    \label{RndAvgISlow}
    & = & \int d \omega'
        \left(
        \frac{e^{-\omega'^2/ (2 \lambda_{\varphi}^2) } }
            {\sqrt{2 \pi \lambda_{\varphi}^2 }} \right)
            \langle I \rangle \Big|_{\ve_d \ra \ve_d - \omega'}
\end{eqnarray}
In the ``slow" limit, $\langle I \rangle$ is inhomogeneously
broadened--  while the random potential $\eta(t)$ is essentially
constant for each event where an electron tunnels on and then off
the level, it does fluctuates from event to event. The effect of
$\eta$ on $\langle I \rangle$ can thus be mimicked by simply
averaging over a Gaussian distribution of level positions $\ve_d$.
As a result, the effect of dephasing is not simple lifetime
broadening, and the conductance lineshape is manifestly
non-Lorentzian; this is in contrast to the voltage probe models.
Nonetheless, if we associate $\lambda_{\varphi}$ with
$\Gamma_{\varphi}$, the height and width of the resonance
described by Eq. (\ref{RndAvgISlow}) for $\langle I \rangle$ is
similar to that obtained in the voltage probe model.

Turning to the noise for the ``slow" random potential model, we
remark again that we focus throughout this section on the
intrinsic contribution to the noise defined by the first term in
Eq. (\ref{TwoNoises}). We first calculate the phase correlation
factor using Eq. (\ref{AvgK2}):
\begin{equation}
    C(t;\tau_1,\tau_2) =
        \exp\left[
            \lambda_{\varphi}^2  \tau_1 \tau_2
            \right]
            \label{SlowC}
\end{equation}
Again, $C(t;\tau_1,\tau_2)$ describes the correlations in the
random phases acquired by two tunnel events (dwell times $\tau_1,
\tau_2$) with a time separation $t$.  Here,  $C(t;\tau_1,\tau_2)$
is independent of time, implying (as expected) only elastic
contributions to $S_Q$ and $S_I$. However, the corresponding fact
that phase correlations persist even at large time separations
(i.e. $C(t \ra \infty) \neq 1$) implies that the elastic
contribution to $S_Q$ is {\it not} given by the simple result of
Eq. (\ref{RndSQNoC}).  It follows that the charge noise in this
limit cannot agree with the intrinsic charge noise $[S_Q]_{bare}$
(Eq. (\ref{DephVPSQ})) found in the dephasing voltage probe model.

One can simply relate the diffuson and interference contributions
to the $\eta$-averaged Green function, resulting in:
\begin{eqnarray}
    S_Q  & = &
                \frac{4 e^2}{\pi \Gamma_0}
                \int d \omega \Big[
                    \bar{f}(\omega) (1-\bar{f}(\omega))
                    \nonumber \\
        &&
                \left(1 - \Gamma_0 \frac{\partial}{\partial \Gamma_0} \right)
                \langle - \textrm{Im } G^R(\omega) \rangle_\eta
                \Big]
                \label{SlowSQ}
\end{eqnarray}
The two terms in this expression correspond to the diffuson and
interference contributions respectively.  As can easily be
verified, Eq. (\ref{SlowSQ}) is simply the charge noise of the
fully coherent system (i.e. Eq. (\ref{RndSQNoC}) at $\eta=0$)
averaged over a Gaussian distribution of level positions; this is
analogous to what was found for the average current.  In the
strong dephasing limit, $\langle G^R \rangle_\eta$ becomes
independent of $\Gamma_0$, and we thus have at zero temperature
(but arbitrary voltage):
\begin{equation}
    S_Q = 4 \frac{ | \langle I \rangle |}{(\Gamma_0)^2}
    \label{FinalSlowSQ}
\end{equation}
This expression is {\it identical} to what is found in both the
inelastic and elastic voltage probe models in the strong dephasing
limit (c.f. Eq. (\ref{VPSQLargeDeph})), and for small voltages,
scales as  $eV / (\lambda_{\varphi} \Gamma_0)$.  The
correspondence to these voltage probe models is not surprising, as
in the strong dephasing limit, the charge noise in these models is
dominated by the fluctuations of the potential in the voltage
probe.  Not surprisingly, these fluctuations are equivalent to
exposing the level directly to a slow, random potential.

Turning to the current noise $S_I$, we find at zero temperature:
\begin{eqnarray}
    S_I & = & \frac{2 e^2 \Gamma_L \Gamma_R}{\pi \Gamma_0}
            \int_{-\infty}^{\infty} d \bar{\omega}
                f_{L}(\bar{\omega} )
            \left(1-
                f_{R}(\bar{\omega})
            \right)  \nonumber \\
    &&
        \left[
            \frac{
                    \Gamma_L^2 +  \Gamma_R^2  }
            {\Gamma_0^2}
                 +
          \frac{2 \Gamma_L \Gamma_R }{\Gamma_0}
            \frac{\partial}{\partial \Gamma_0}
            \right]
            \langle - \textrm{ Im} G^R(\omega) \rangle_\eta
\end{eqnarray}
Taking the large dephasing limit ($\lambda_{\varphi} \gg
\Gamma_0$) yields:
\begin{eqnarray}
    S_I & \ra &
        2 e \left(
                \frac{ \Gamma_L^2 + \Gamma_R^2 }{ \Gamma_0^2}
        \right)
        \langle I \rangle
\end{eqnarray}
The Fano factor here has the familiar form corresponding to the
shot noise of two classical Poisson processes in series, and
agrees with the voltage probe result (c.f. Eq. (\ref{IncFano})).
Thus, we find that the slow random potential model and the
dephasing voltage probe model {\it agree} both on the form of
$S_I$ and $S_Q$ in the strong dephasing limit; this is despite the
fact that they yield quite different conductance lineshapes.

Turning to the question of the quantum limit, shown in Figs.
\ref{QLPlot4} and \ref{QLPlot5} is $\chi$ as function of dephasing
strength for the ``slow" random potential model at zero
temperature and small voltage.  The results obtained for $S_I$ and
$S_Q$ suggest that at large dephasing strengths, $\chi$ in the
slow potential model will be the same as that found in the
dephasing voltage probe model.  As seen in Figs. 4 and 5, we find
that a reasonable correspondence holds even at moderate levels of
dephasing . This does not however imply that the two models are
always equilvaent. We have also calculated $\chi$ in the large
voltage, ``co-tunneling" limit considered by Averin
\cite{AverinRLM}, see Fig. \ref{QLPlotLargeV}.  Here, there is a
marked quantitative difference between the slow potential model
and the dephasing voltage probe model, with $\chi$ being
suppressed more quickly by dephasing in the latter model.  Again,
we see that both the nature of the dephasing source as well as the
strength of dephasing influences the suppression of $\chi$.

The departure from the quantum limit in the fluctuating random
potential model may be understood as it was in the dephasing
voltage probe model (see text following Eq.
(\ref{VPSQLargeDeph}))-- there is a classical uncertainty in the
system stemming from the source of dephasing; this uncertainty in
turn leads to extraneous noise in both $I$ and $Q$ which takes one
away from the quantum limit.

Finally, it is instructive to look at the results of this section
while retaining a small bandwidth $\Omega$ for the spectral
density $J(\omega)$.  This will give further insight on the
contribution $[S_Q]_{class}$ (c.f. Eq. (\ref{TwoSQ})) to the
charge noise in the voltage probe models, a contribution arising
from fluctuations of the probe voltage.
At small but finite bandwidth $\Omega$, the phase correlation
factor will be given to a good approximation by:
\begin{equation}
    C(t;\tau_1,\tau_2) = \exp\left[
        \lambda_{\varphi}^2 \tau_1 \tau_2 e^{-\Omega |t|}
        \right]
\end{equation}
For long times $|t| \gg 1/{\Omega}$, phase correlations will
vanish, and the elastic contribution to $S_Q$ will again be given
by Eq. (\ref{RndSQNoC}) as opposed to Eq. (\ref{SlowSQ}). However,
the presence of a small but finite bandwidth implies that there
will now also exist an inelastic contribution to $S_Q$ (c.f. Eq.
(\ref{RndSQInelast})), corresponding to small energy transfers
$\Delta \omega \sim \Omega$. The quasi-probability $P(\Delta
\omega;\tau_1,\tau_2)$ for having an inelastic event will be given
by (c.f. Eq. (\ref{PEDefn})):
\begin{eqnarray}
    P_{D/I}(\Delta \omega;\tau_1,\tau_2)  =
    \frac{1}{\pi} \sum_{n=1}^{\infty} \frac{
        \left( \lambda_{\varphi} \tau_1 \tau_2 \right)^n}{n!}
        \frac{ n \Omega}{ (\Delta \omega)^2 + (n \Omega)^2}
    \label{SlowPE}
\end{eqnarray}
For typical dwell times $\lambda_{\varphi}^2 \tau_1 \tau_2 $ is
order unity, and thus the only significant contribution to the sum
in Eq. (\ref{SlowPE}) will come from the first few terms. Further,
note that in Eq. (\ref{RndSQInelast}) for the inelastic
contribution to $S_Q$, $\Delta \omega$ appears only in the
distribution functions of the leads. Thus, since by assumption we
have $e V, \Gamma_0 \gg \Omega$, we may safely replace the
Lorentzians in the above expression by delta functions, and the
inelastic contribution may be accurately represented by an elastic
one:
\begin{equation}
    P_{D/I}(\Delta \omega;\tau_1,\tau_2) \ra \delta(\Delta \omega) \left(
        C(t=0;\tau_1,\tau_2)^{\pm 1} - 1 \right)
\end{equation}
In the case of $S_Q$, combining this ``quasi-elastic" contribution
with the pure elastic contribution given by Eq. (\ref{RndSQNoC})
results again in Eq. (\ref{SlowSQ}), which was found above by
taking the $\Omega \ra 0$ limit from the outset.  We thus see that
the slow potential model includes the contribution from inelastic
events involving a small energy transfer $\sim \Omega$; keeping
only purely elastic contributions would yield the result of Eq.
(\ref{RndSQNoC}) for $S_Q$.  As already noted, the purely elastic
contribution coincides with the intrinsic contribution to the
charge noise ($[S_Q]_{bare}$) found in the dephasing voltage probe
model. It thus follows that the ``quasi-elastic" contribution in
the random potential model (i.e. stemming from Eq. (\ref{SlowPE}))
corresponds to the contribution $[S_Q]_{class}$ in the voltage
probe model (i.e. charge fluctuations induced by potential
fluctuations in the probe, c.f. Eq. (\ref{DephVPSQClass})).

\subsection{Results from the fast random potential model}

The second limiting case of the random potential model is that of
a ``fast" random potential, where $J(\omega)$ is flat on the
frequency scales of interest. This will allow us to make the
replacement:
\begin{equation}
    J(\omega) \ra J(\omega = 0)
    \label{FastApprox}
\end{equation}
A necessary requirement for the ``fast" limit is that the
characteristic bandwidth of the random potential $\Omega$ is much
larger than the frequency scales of interest (i.e. reverse the
inequality in Eq. (\ref{OmegaBound})).  Given this large bandwidth
$\Omega$, we expect that inelastic contributions to $S_Q$ and
$S_I$ will be important; this makes it unlikely that the total
current or charge noise in this limit will agree with either of
the voltage probe models.  In what follows, we will consider
separately the elastic and inelastic contributions to the noise.

Turning first to the average current in the presence of a ``fast"
random potential, Eq. (\ref{RndAvgG}) yields:
\begin{equation}
    \langle G^R(\tau) \rangle =
        e^{-i \ve_d \tau} e^{- (\Gamma_0 + \Gamma_{\varphi}) \tau / 2}
    \label{RndAvgGFast}
\end{equation}
where
\begin{equation}
    \Gamma_{\varphi} =  J(\omega \ra 0)
    \label{RndGammaPhi}
\end{equation}
Thus, for the average Green function and average current, the
effect of a ``fast" random potential is just a simple lifetime
broadening, identical to the situation in the voltage probe
models.  As a result, Eq. (\ref{RndAvgI2}) yields the same
Lorentzian result for $\langle I \rangle$ as the dephasing voltage
probe model, Eq. (\ref{DephVPI}).  Note that for strong dephasing
($\Gamma_{\varphi} \gg \Gamma_0$), the approximation of Eq.
(\ref{FastApprox}) is valid only if $\Omega \gg \Gamma_{\varphi}$
in addition to the reversed-ineqality of Eq. (\ref{OmegaBound}).
If this is not true, the contribution from short times in Eq.
(\ref{RndAvgI2}) will dominate, and one will not obtain a
Lorentzian form for the average current (instead, the results for
a ``slow" random potential, discussed in the previous subsection,
will apply).

We turn next to the calculation of $S_Q$ and $S_I$ for a ``fast"
random potential.  We find that the factor $C(t;\tau_1,\tau_2)$
(c.f. Eq. (\ref{CFactor})) describing the correlation between the
random phases of a pair of tunnel events is given by:
\begin{equation}
    C(t;\tau_1,\tau_2) =
        \exp\left[
             \Gamma_{\varphi} \tau_{overlap}
            \right]
\end{equation}
where $\tau_{overlap}$ is the overlap between the two time
intervals describing the tunneling events (i.e. $[t-\tau_1,t]$ and
$[0-\tau_2,0]$), and $\Gamma_{\varphi}$ is defined in Eq.
(\ref{RndGammaPhi}).  As was anticipated in the discussion prior
to Eq. (\ref{RndSQNoC}), correlations in phase between the two
tunnel events are important only if the time separation $t$ is
sufficiently small. The correlations vanish as $t \ra \infty$
(i.e. $C \ra 1$), and thus the elastic contribution to $S_Q$ is
given by Eq. (\ref{RndSQNoC}).  As with the result for $\langle I
\rangle$, this expression agrees exactly with the dephasing
voltage probe result of Eq. (\ref{DephVPSQ}). However, in sharp
contrast, the elastic contribution to $S_I$ deviates strongly from
the voltage probe model. At $T=0$, we find for the elastic
contribution:
\begin{eqnarray}
    S_I \big|_{elast}  =  \frac{2 e^2}{2 \pi}
        \Gamma_L \Gamma_R
        \int_{\mu_R}^{\mu_L} d \omega
            \left| \langle G^R(\omega) \rangle \right|^4
            \left[
                 \omega^2 + \Gamma^2
                \left(
                    \frac{\Gamma_L-\Gamma_R}{2 \Gamma_0}
                \right)^2
            \right]
\end{eqnarray}
Taking the strong dephasing limit $\Gamma_{\varphi} \gg \Gamma_0$,
this yields a vanishing Fano factor:
\begin{equation}
    f_{elast} \equiv \frac{ S_I \big|_{elast} }{2 e \langle I
    \rangle}  \propto \frac{ \Gamma_0}{\Gamma_{\varphi}}
\end{equation}
In contrast, the voltage probe models yield a non-vanishing Fano
factor in the incoherent limit, given by the classical expression
of Eq. (\ref{IncFano}).

We turn now to the inelastic contributions to $S_I$ and $S_Q$
which are non-vanishing even at zero temperature and zero voltage.
These inelastic contributions always {\it increase} the noise in
the fluctuating potential model, while the opposite is found in
voltage probe models:  the inelastic version of the voltage probe
model yields {\it smaller} values of $S_Q$ and $S_I$ than the
purely elastic version.  The inelastic contributions will be
determined by the quasi-probabilities $P_{D/I}(\Delta
\omega;\tau_1,\tau_2)$ defined in Eq. (\ref{PEDefn}).  We find
\begin{widetext}
\begin{eqnarray}
        P_{D/I}(\Delta \omega;\tau_1,\tau_2)
     & = &
        \frac{1}{\pi} \frac{\Gamma_{\varphi}}{(\Delta \omega)^2 +
        \Gamma_{\varphi}^2} \Bigg[
            \frac{\Gamma_{\varphi}}
                { \Delta \omega }
            \left(
                e^{\pm \Gamma_{\varphi} \tau_<}
                \sin \left( \frac{\Delta \omega (\tau_> -
                \tau_<)}{2} \right) -
                \sin \left( \frac{\Delta \omega (\tau_> +
                \tau_<)}{2} \right)
            \right)
        \nonumber \\
    &&
            \pm
                \left(
                      e^{\pm \Gamma_{\varphi} \tau_<}
                      \cos \left( \frac{\Delta \omega (\tau_> -
                      \tau_<)}{2} \right) -
                      \cos \left( \frac{\Delta \omega (\tau_> +
                      \tau_<)}{2} \right)
                \right) \Bigg]
    \label{InelasticKernel}
\end{eqnarray}
\end{widetext}
Here, $\tau_>$ ($\tau_<$) is the greater (lesser) of
$\tau_1,\tau_2$.  As expected, inelastic processes involving
energy transfers $\Delta \omega \simeq \Gamma_{\varphi}$ can make
a sizeable contribution to $S_Q$.  In the strong dephasing limit
($\Gamma_{\varphi} \gg \Gamma_0$), these inelastic processes are
an unavoidable consequence of having a Lorentzian form for the
average current, as this requires a noise bandwidth $\Omega \gg
\Gamma_{\varphi}$.  In the strong dephasing limit, we find the
simple result:
\begin{eqnarray}
    S_Q |_{inelastic} & = & \frac{2 e^2}{\pi^2}
        \frac{1}{\Gamma_0}
        \int d \omega_1 \int d \omega_2
            \bar{f}(\omega_1) (1 - \bar{f}(\omega_2) )
        \nonumber \\
        &&
            \langle \textrm{Im } G^R(\omega_1) \rangle_\eta
            \langle \textrm{Im } G^R(\omega_2) \rangle_\eta
        \label{RndSQInelastic}
\end{eqnarray}
In the small voltage, large dephasing limit, the inelastic
contribution to $S_Q$ scales as $(1/ \Gamma_0)$, and is much
larger than the elastic contribution, which scales as $e V /
\Gamma_{\varphi}^2$.

Similarly for $S_I$, we find an inelastic contribution in the $V
\ra 0$ limit given by:
\begin{eqnarray}
     S_I \big|_{inelast}  & = &  \frac{2 e^2}{2 \pi}
        \frac{\Gamma_L \Gamma_R}{\Gamma_0}
        \int d \omega_1 \int d \omega_2 \nonumber \\
    &&
            \left[
                f(\omega_1)(1-f(\omega_2)) +
                (1-2 f(\omega_1)) \frac{d f(\omega_2)}{d \omega} eV
            \right]
    \nonumber \\
    &&
            \times \big \langle \textrm{Im } G^R(\omega_1) \big \rangle
            \big \langle \textrm{Im } G^R(\omega_2) \big \rangle
\end{eqnarray}
Unlike the elastic contribution, this term does not vanish in the
strong dephasing limit.

\begin{table*}
\caption{\label{SynopsisTable} Synopsis of results for the various
dephasing models, for the limit of small voltage ($eV \ll \Gamma$)
and large dephasing ($\Gamma_0 / \Gamma \ra 0$).  The classical
Fano factor $f_{classical}$ appearing in the third column is
defined in Eq. (\ref{IncFano}).  For the inelastic and dephasing
voltage probe models, we distinguish the intrinsic contribution to
$S_Q$ from that arising from fluctuations of the probe voltage.}
\begin{ruledtabular}
\begin{tabular}{ccccc}
    Dephasing Source &
    $\langle  d I / d V \rangle$ &
    $S_I = 2 e f \langle I \rangle$ &
    $S_Q$ &
    $\chi$ \\
\hline
\hline
    ``pure escape" voltage probe &
    Lorentzian &
    $\frac{1}{2} \leq f \leq 1$ &
    $\propto \frac{\Gamma_0}{\Gamma} \ra 0$ &
    $\frac{ 1}{2 f} \frac{(\mu-\ve_d)^2}{(\mu-\ve_d)^2 + \Gamma^2/4}$
    \\
\hline
    Inelastic voltage probe &
    Lorentzian &
    $f \ra f_{classical}$ &
    $[S_Q]_{bare} \propto \frac{\Gamma_0}{\Gamma} \ra 0$ &
    $\propto \left( \frac{\Gamma_0}{\Gamma}\right)^2  \ra 0 $ \\
    & & &
    $[S_Q]_{class} \ra \frac{4 \langle I \rangle }{(\Gamma_0)^2} \propto  \frac{\Gamma }{
        \Gamma_0 }$ &
    \\
\hline
    Dephasing voltage probe &
    Lorentzian &
    $f \ra f_{classical}$ &
    $[S_Q]_{bare}$ independent of $\frac{\Gamma_0}{\Gamma}$ &
    $\propto \left( \frac{\Gamma_0}{\Gamma}\right)^2  \ra 0 $ \\
    & & &
    $[S_Q]_{class} \ra \frac{4\langle I \rangle }{(\Gamma_0)^2} \propto  \frac{\Gamma }{
        \Gamma_0 }$ &
    \\
\hline
    ``Slow" fluctuating potential &
    Voit profile (c.f. Eq. \ref{RndAvgISlow}) &
    $ f \ra f_{classical}$ &
    $\frac{4 \langle I \rangle }{(\Gamma_0)^2} \propto  \frac{\Gamma }{ \Gamma_0 }$ &
    $\propto \left( \frac{\Gamma_0}{\Gamma}\right)^2  \ra 0 $ \\
\hline
    ``Fast" fluctuating potential &
    Lorentzian &
    voltage independent &
    voltage independent &
    $\propto \left(\frac{eV \Gamma_0}{\Gamma^2}\right)^2 \ra 0$ \\
\end{tabular}
\end{ruledtabular}
\end{table*}

Finally, we turn to the issue of the quantum limit for the ``fast"
fluctuating potential model, in the small voltage limit. If one
(rather unphysically) only retains the elastic contributions, we
have that $\chi \sim 1$ even in the strong dephasing limit.  This
is due to the strong suppression of $S_I \big|_{elast}$.  Keeping
the inelastic terms, one finds instead that $\chi$ is greatly
suppressed. In the strong dephasing limit, we have:
\begin{equation}
    \chi \propto \left( \frac{eV}{\Gamma }\right)^2
        \left( \frac{\Gamma_0}{\Gamma}  \right)^2
\end{equation}

\section{Conclusions}

We have studied the noise properties and detector efficiency of a
mesoscopic resonant-level system subject to dephasing.  We find
that these properties are sensitive both to the degree of detector
coherence (i.e. the ratio $\Gamma_0 / \Gamma$) {\it and} to the
nature of the dephasing source.  This is in contrast to the
average current, which is largely insensitive to the degree of
coherence. Thus, even though one may have a well defined resonance
in the average current, its suitability for use as a detector will
depend strongly on its coherence properties.

We find that in general, different models of dephasing all lead to
a suppression of the measurement efficiency $\chi$, though the
rate at which this occurs depends on the model (see Figs. 1-5).
The only exception to the above is dephasing arising from the
``pure escape" voltage probe model. Here, $\chi$ may remain order
unity even in the strong dephasing limit, a limit in which all
transport involves entering the voltage probe and then leaving it
incoherently. Though this result appears surprising, the ``pure
escape" model is unique among those considered, as there is no
lost information associated with the dephasing source, nor is
there any classical contribution to the detector noise arising
from the detector being in a mixed (i.e. non-pure) state; this is
why it is able to remain quantum limited in the fully incoherent
limit.  Among the more conventional voltage probe models, we find
that $\chi$ is suppressed more quickly to zero by dephasing in the
dephasing voltage probe model than in the inelastic voltage probe
model (see Figs. 1 and 2).  The differences between the models
considered is summarized in Table 1.

Although the models considered here for dephasing may be regarded
as describing to some extent the effect of interactions on the
measurement efficiency of quantum detectors, it will be
interesting to study this question in more general setups and in
models in which strong interactions are directly included.

We thank S. Girvin and F. Marquadt for useful discussions, and M.
Buttiker for a critical reading of the original manuscript and
pointing out the importance of the second term in Eq.
(\ref{TwoSQ}). This work was supported by the Keck foundation, and
by the NSF under grant No. DMR-0084501.

\end{document}